\documentclass[prc,aps,nofootinbib,twocolumn,superscriptaddress]{revtex4-2} 
\usepackage{graphicx} 
\usepackage[dvipdfmx]{epsfig} 
\usepackage{bm}
\usepackage[normalem]{ulem}
\usepackage{amsmath,color}

\renewcommand{\vec}[1]{\mbox{\boldmath $#1$}}
\newcommand{\trento}{T\raisebox{-.5ex}{R}ENTo~}

\begin{document}

\title{Orientation effects of deformed nuclei in intermediate-energy nuclear scattering}
\author{W. Horiuchi}
 \email{whoriuchi@omu.ac.jp}
\affiliation{Department of Physics, Osaka Metropolitan University, Osaka 558-8585, Japan}
\affiliation{Nambu Yoichiro Institute of Theoretical and Experimental Physics (NITEP), Osaka Metropolitan University, Osaka 558-8585, Japan}
\affiliation{RIKEN Nishina Center, Wako 351-0198, Japan}

\author{K. Hagino}
\email{hagino.kouichi.5m@kyoto-u.ac.jp}
\affiliation{ 
Department of Physics, Kyoto University, Kyoto 606-8502, Japan} 
\affiliation{Institute for Liberal Arts and Sciences, Kyoto University, Kyoto 606-8501, Japan}
\affiliation{RIKEN Nishina Center, Wako 351-0198, Japan}

\begin{abstract}
  We investigate the effects of nuclear deformation on intermediate-energy
  nuclear scattering within the Glauber theory.
  A formalism that explicitly incorporates the orientation dependence
  of deformed intrinsic states is derived, 
  in which reaction observables are first computed with a fixed orientation and then the angle average of the reaction observables is taken afterwards. This formalism is 
  applied to elastic scattering
  and total reaction cross sections. In addition, a new expansion scheme
  of the optical-limit approximation is proposed and shown to include
  contributions from a class of nuclear excitations, allowing a simple
  estimate of part of the total inelastic scattering cross section.
  The orientation dependence is generally weak for
  both proton- and carbon-target scattering
  and supports the validity of conventional treatments with an angle averaged density
  for light and medium-heavy nuclei. However, noticeable deviations emerge
  for strongly deformed heavy nuclei. The present analysis further demonstrates
  that the summed inelastic scattering cross section to rotationally
  excited states can be interpreted as a manifestation of fluctuations
  of the nuclear surface density associated with deformation.
  The present study provides a quantum-mechanical interpretation
  of deformation effects in intermediate-energy nuclear reactions
  and suggests a pathway toward a more reliable theoretical description
  of interaction cross section measurements. Brief comments on the role of nuclear deformation in 
  low-energy heavy-ion reactions and ultra-relativistic heavy-ion collisions are also made.
\end{abstract}

\maketitle

\section{Introduction}

Nuclear deformation is one of the most fundamental concepts
in nuclear structure physics \cite{BohrMottelson2,RingSchuck,Greiner}. The remarkable success of 
the 
collective model of Bohr and Mottelson has established the picture that many nuclei exhibit
intrinsically deformed shapes, giving rise to characteristic rotational
spectra and enhanced electromagnetic transition strengths.
However, the notion of nuclear rotation is, to some extent, a classical concept,
and its interpretation in a finite quantum many-body system requires careful
consideration.

For even-even nuclei, the ground state has spin and parity of $J^\pi=0^+$,
implying that the nucleon distribution is isotropic in the laboratory frame.
A deformed intrinsic state, if it exists, must therefore be represented
as a superposition of wave functions with all possible orientations
with equal weights.
It should be emphasized that this restoration of rotational symmetry 
is fundamentally different from a classical rotating object.
Consequently, the question of how nuclear deformation manifests
itself in reaction observables remains nontrivial.

Traditionally, nuclear deformation has been investigated
through low-energy observables such as excitation energies of the first $2^+$
states, electric quadrupole transition strengths $B(E2)$, and quadrupole moments~\cite{BohrMottelson2,Raman01}.
More recently, interaction-cross-section and total-reaction-cross-section
measurements at intermediate energies have emerged as powerful probes
of nuclear size and deformation~\cite{Minomo11,Sumi12,Horiuchi12,Horiuchi22}.
In many cases, deformation leads to an increase
in the root-mean-square radius and consequently enhances
the total reaction or interaction cross section.
Such measurements have therefore been extensively used
to study the deformation of stable and unstable nuclei.

Despite these developments, the interpretation of deformation effects
in intermediate-energy nuclear reactions remains an open issue.
In particular, there is no consensus on whether deformation
should be treated as a coherent superposition of static intrinsic
configurations or as a rotational motion during the collision process.
This question is especially relevant at intermediate and high energies,
where the energy scale for the collision is much larger than that for the rotational motion. 
In the classical view, this may be translated into the difference of the time scales. 
For incident energies of approximately 100--1000 MeV per nucleon,
the characteristic time required for a projectile nucleus to traverse
a target nucleus is on the order of 10$^{-23}$--$10^{-22}$~s,
which is much shorter than typical rotational timescales
$\sim 10^{-21}$--$10^{-20}$~s of deformed nuclei.

The difference in the energy and the time scales is even larger in ultra-relativistic heavy-ion collisions. 
This observation has led to the viewpoint that ultra-relativistic-energy reactions
effectively probe event-by-event snapshots of nuclear configurations~\cite{giacalone2020,giacalone2021,jia2023,zhang2022,ryssens2023,jia2024,star2024,bally2022,jia2022,jia2022-2,giacalone2025,duguet2025,dimri2023,hagino2025}.
Based on this idea, 
in recent years, studies of nuclear deformation based on high-energy scattering
experiments have been increasingly active. However, the relationship
between experimentally extracted reaction observables
and the underlying quantum-mechanical picture of deformation
has not yet been fully clarified \cite{doba2025}. A consistent framework is therefore
required to understand what aspects of nuclear shape are actually
observed in high-energy reactions.

The theoretical description of deformed nuclei
in intermediate-energy nuclear scattering has already been discussed in Refs. \cite{glauber1990,Tostevin1999}. 
In this work, we revisit this problem and 
propose a quantum-mechanical interpretation
of deformation effects in such reactions, see also Refs. \cite{ke2025,xu2026} for related works. 
Particular attention is paid to the role of orientation dependence
in deformed nuclei. Using the Glauber theory, we systematically investigate
a variety of nuclei ranging from light to heavy deformed nuclei. 
By comparing calculations that explicitly
include orientation effects with those based on orientation-averaged densities,
we analyze their impact on elastic-scattering and total reaction cross sections.
We also examine the quasielastic scattering cross section,
which provides a particularly sensitive probe of nuclear deformation.
Through these investigations, we aim to clarify how nuclear deformation
manifests itself in high-energy reaction observables
and what information on nuclear shape
can be extracted from scattering experiments.

This paper is organized as follows.
In Sec.~\ref{formalism.sec}, we present the theoretical framework.
Section~\ref{angular.sec}
briefly reviews the angular-momentum projection method used
to construct rotational states from a deformed intrinsic state.
In Sec.~\ref{cs.sec}, the formulas incopolating explicit deformation
for elastic scattering and total reaction
cross sections within the Glauber theory are derived.
Section~\ref{qe.sec} discusses the quasielastic scattering cross section,
while Sec.~\ref{ola.sec} presents the optical-limit approximation
and its relation to density fluctuations. Numerical results
are presented in Sec.~\ref{results.sec}. After describing the model setup
in Sec.~\ref{model.sec}, elastic, reaction, and quasielastic cross sections
are discussed in Sec.~\ref{csres.sec}. 
In Secs. ~\ref{fusion.sec} and ~\ref{relhic.sec}, 
we also discuss relations to the 
treatments of nuclear deformation in low-energy heavy-ion reactions and ultra-relativistic heavy-ion 
collisions, respectively. 
Finally, the conclusions of this work are summarized in Sec.~\ref{conclusion.sec}.

\section{Formalism}
\label{formalism.sec}

\subsection{Angular momentum projection of the intrinsic wave function}
\label{angular.sec}

We consider a nuclear state with total angular momentum $J$ and its projection $M$, constructed by performing angular-momentum and parity projections on a single intrinsic wave function $\left|\Phi\right>$:
\begin{align}
  \left|\Psi_{JM}^\pi\right>=\sum_K g_K\hat{P}_{MK}^{J}P^\pi\left|\Phi\right>,
\end{align}
with the angular-momentum projection operator
\begin{align}
  \hat{P}_{MK}^{J}=\frac{2J+1}{8\pi^2}\int d\bm{\omega} D_{MK}^{J*}(\bm{\omega})\hat{R}(\bm{\omega}).
\end{align}
Here, $\hat{R}(\bm{\omega})$ denotes the rotation operator specified by the Euler angles $\bm{\omega}$.
For an axially symmetric intrinsic state in an even-even nucleus,
only the $K=0$ component contributes,
and the parity is necessarily positive.
The wave function then reduces to
\begin{align}
  \left|\Psi_{JM}\right>=g_0\hat{P}_{M0}^J\left|\Phi\right>,
\end{align}
with the normalization factor given by
\begin{align}
  |g_0|^2=\frac{1}{\left<\Phi\right|\hat{P}_{00}^J\left|\Phi\right>}
\end{align}
The Wigner $D$ function for $K=0$ can be expressed as
\begin{align}
  D_{M0}^{J*}(\omega)=\sqrt\frac{4\pi}{2J+1}Y_{JM}({\bm{\Omega}}),
\end{align}
which depends only on the orientation angle ${\bm{\Omega}}=(\theta,\phi)$.
Accordingly, the angular-momentum projection operator for $K=0$ becomes
\begin{align}
  \hat{P}_{M0}^{J}=\sqrt\frac{2J+1}{4\pi}\int d{\bm{\Omega}}\,Y_{JM}({\bm{\Omega}})\hat{R}({\bm{\Omega}}),
\end{align}
where $d{\bm{\Omega}}=\sin\theta d\theta d\phi$.

\subsection{Elastic and total reaction cross sections in the Glauber formalism}
\label{cs.sec}

Let us first consider elastic scattering of a deformed projectile  
nucleus $\Psi_0^P$ off a spherical target nucleus $\Psi_0^T$. 
For simplicity, the Coulomb interaction is omitted in the following formulation. It can be incorporated following the procedure described in Refs.~\cite{Suzuki03,Horiuchi26a,Horiuchi26b}.
According to the Glauber theory, the elastic-scattering amplitude is given by 
\begin{align} f(\bm{q})=\frac{ik}{2\pi}\int d\bm{b}\,e^{-i\bm{q}\cdot\bm{b}} 
\left<\Psi_0^P\Psi^T_0 \right|1-e^{i\chi(\bm{b})} \left|\Psi_0^P\Psi^T_0\right>, 
\label{elamp.eq} 
\end{align} 
with the Glauber phase-shift function 
\begin{align} \chi(\bm{b})=\sum_{i\in P}\sum_{j\in T}\chi_{NN}(\bm{b}+\hat{\bm{s}}_i^P-\hat{\bm{s}}_j^T). 
\end{align} 
Here, $\bm{b}$ is the impact-parameter vector, and $\hat{\bm{s}}_i^P$ ($\hat{\bm{s}}_i^T$) 
denotes the transverse-coordinate operator of the $i$th nucleon in the projectile (target) nucleus, perpendicular to the beam direction. The two-dimensional momentum-transfer vector $\bm{q}$ has the magnitude $q=2k\sin(\theta_{\rm cm}/2)$, where $k$ is the relativistic wave number for the relative motion between the colliding nuclei and $\theta_{\rm cm}$ is the scattering angle in the center-of-mass frame.

For simplicity, we consider a well-deformed intrinsic state
for the projectile nucleus and restrict ourselves to the ground state
with angular momentum $J=0$.
The projectile wave function is written as
\begin{align}
  \left|\Psi_0^P\right>=n_0\hat{P}_{00}^0\left|\Phi^P\right>
  \end{align}
with $|n_0|^2=\left<\Phi^P\right|\hat{P}_{00}^0\left|\Phi^P\right>^{-1}$, where
\begin{align}
  \hat{P}_{00}^0=\frac{1}{4\pi}\int d{\bm{\Omega}}\,\hat{R}({\bm{\Omega}}).
\end{align}
Using the projected wave function,
the scattering amplitude in Eq.~(\ref{elamp.eq}) can be expressed as
\begin{widetext}
\begin{align}
  f(\bm{q})&=\frac{ik}{2\pi}\int d\bm{b}\,e^{-i\bm{q}\cdot\bm{b}} |n_0|^2
  \left(\frac{1}{4\pi}\right)^2\iint d{\bm{\Omega}}d{\bm{\Omega}}'\frac{\left<\Phi^P(\bm{\Omega})\Psi^T_0\right|
    1-e^{i\chi} \left|\Phi^P({\bm{\Omega}')}\Psi^T_0\right>}
       {\left.\left<\Phi^P(\bm{\Omega})\right| \Phi^P(\bm{\Omega}')\right>}
       \left.\left<\Phi^P(\bm{\Omega})\right| \Phi^P(\bm{\Omega}')\right>
\end{align}
\end{widetext}
with $\left|\Phi^P(\bm{\Omega})\right>\equiv \hat{R}({\bm{\Omega}})\left|\Phi^P\right>$. 
For strongly deformed intrinsic states,
the overlap between different orientations becomes sharply peaked around
${\bm{\Omega}}=\bm{\Omega}'$.
By adopting the so-called needle approximation \cite{RingSchuck,duguet2026}, that is,
$\left.\left<\Phi^P(\bm{\Omega})\right| \Phi^P(\bm{\Omega}')\right>
\approx\delta({\bm{\Omega}}-{\bm{\Omega}}')$,
the above expression is reduced to
\begin{align}
  f(\bm{q})\approx\frac{1}{4\pi}\int d\bm{\Omega}\,
  f(\bm{q}; \bm{\Omega})
  \equiv \left<f(\bm{q}; \bm{\Omega})\right>_{\bm{\Omega}}
\end{align}
with the oriantation-dependent scattering amplitude
\begin{align}
  f(\bm{q}; \bm{\Omega})\equiv
  \frac{ik}{2\pi}\int d\bm{b}\,e^{-i\bm{q}\cdot\bm{b}}
  \left(1-\mathcal{S}(\bm{b};\bm{\Omega})\right),
\label{elampaprx.eq}
\end{align}
and $S$-matrix:
\begin{align}
  \mathcal{S}(\bm{b};\bm{\Omega})\equiv
  \left<\Phi^P(\bm{\Omega})\Psi^T_0\right|e^{i\chi(\bm{b})}
  \left|\Phi^P(\bm{\Omega})\Psi^T_0\right>,
\end{align}
The corresponding differential cross section is then given by
\begin{align}
  \dfrac{d\sigma_{\rm el}}{d\Omega_{\rm cm}}=|f(\bm{q})|^2
  \approx
\left|\left<f(\bm{q}; \bm{\Omega})\right>_{\bm{\Omega}}\right|^2.
\label{eldcsaprx.eq}
\end{align}

Following the same procedure, the expression of
the total reaction cross section
\begin{align}
  \sigma_R=\sigma_{\rm tot}-\sigma_{\rm el}
  =\int d\bm{b}\,\left(1-\left|\left<\Psi_0^P\Psi^T_0
\right|e^{i\chi}
\left|\Psi_0^P\Psi^T_0\right>\right|^2\right)
\end{align}
can also be reduced to
the oriantation avarage of the $S$-matrix as
\begin{align}
\sigma_R\approx\int d\bm{b}\,\left(1-
\left|\left<\mathcal{S}(\bm{b}; \bm{\Omega})\right>_{\bm{\Omega}}\right|^2\right).
\label{rcsaprx.eq}
\end{align}

This expression indicates that the orientation of the wave function
remains unchanged during the reaction process.
This is precisely the frozen-orientation picture underlying
the adiabatic approximation in the Glauber theory. The approximation
is well justified at intermediate and high energies,
where the collision time is much shorter, approximately
one to two orders of magnitude shorter
than the characteristic rotational time scale of the nucleus.
In fact, if one constructs a collective rotational Hamiltonian, it entirely arises 
from the off-diagonal components with 
${\bm{\Omega}}\neq\bm{\Omega}'$ \cite{RingSchuck}. 
Therefore, neglecting the off-diagonal components is equivalent to neglecting the rotational Hamiltonian 
and thus the rotational excitation energies, 
which is nothing but the adiabatic approximation. 

\subsection{Quasielastic scattering cross section}
\label{qe.sec}

By extending the above formalism, the quasielastic cross section
can also be evaluated within the framework of the Glauber theory.
Here, the quasielastic cross section is defined
as the inclusive cross section obtained
by summing over all final rotational states with the angular momentum $L$ and its projection $M$ of the projectile,
including the elastic channel \cite{glauber1990,Hagino2022}. It is given by
\begin{align}
  \sigma_{\rm qel}=\sum_{L,M}\int d\bm{b}\, \left|\delta_{L,0}\delta_{M,0}- \left<\Psi_{LM}^P\Psi^T_0 \right|e^{i\chi} \left|\Psi_0^P\Psi^T_0\right>\right|^2.
\end{align}
The ``interaction'' cross section is defined by
\begin{align}
  \sigma_{I}&\equiv \sigma_{\rm tot}-\sigma_{\rm qel}\notag\\
  &=\int d\bm{b}\,\Big(1-\sum_{L,M}\left|\left<\Psi_{LM}^P\Psi^T_0
\right|e^{i\chi}
\left|\Psi_0^P\Psi^T_0\right>\right|^2\Big).
\end{align}
This expression corresponds directly to the interaction cross section
(or mass-changing cross section) measured in experiments using a proton target,
provided that the states with $LM$ are restricted to particle bound states 
and other excitations to non-rotational bound states can be neglected.
For nucleus-nucleus collisions, on the other hand,
excitations of the target nucleus should also be subtracted.
Since such contributions are not excluded in the present formulation,
the resulting cross section is expected
to yield a somewhat larger cross section than the experimentally
measured interaction cross section.

The rotational states with the angular momentum $L$ and its $z$ component $M$ are 
generated from the intrinsic wave function as 
\begin{align}
  \left|\Psi_{LM}^{P}\right>=n_L\sqrt\frac{2L+1}{4\pi}\int d{\bm{\Omega}}\, Y_{LM}({\bm{\Omega}}) \left|\Phi^P(\bm{\Omega})\right>
\end{align}
with $|n_L|^2=\left<\Phi^P\right|\hat{P}_{00}^L\left|\Phi^P(\bm{\Omega})
\right>^{-1}$.
It should be noted that the so generated states belong
to the rotational band built on the intrinsic state.
Since $K=0$ in the present case,
only even values of the angular momentum $L$ are allowed.
Using this projected wave function, one obtains
\begin{widetext}
\begin{align}
  \left<\Psi_{LM}^P\Psi^T_0 \right|e^{i\chi} \left|\Psi_0^P\Psi^T_0\right>
  &=n_Ln_0\sqrt\frac{2L+1}{4\pi}\frac{1}{\sqrt{4\pi}}
  \iint d{\bm{\Omega}} d{\bm{\Omega}}'\, Y_{LM}^{*}({\bm{\Omega}})Y_{00}({\bm{\Omega}}')
  \notag\\
  &\times
  \frac{  \left<\Phi^P(\bm{\Omega})\Psi^T_0 \right|e^{i\chi} \left|
  \Phi^P(\bm{\Omega}')\Psi^T_0\right>}
{\left.\left<\Phi^P(\bm{\Omega})\right| \Phi^P(\bm{\Omega}')\right>}
       \left.\left<\Phi^P(\bm{\Omega})\right| \Phi^P(\bm{\Omega}')\right>,\\
  &\approx\frac{1}{\sqrt{4\pi}}\int d{\bm{\Omega}}\,Y_{LM}^*({\bm{\Omega}})\mathcal{S}(\bm{b}; \bm{\Omega}), 
\end{align}
\end{widetext}
where the needle approximation has been employed in the final step
with $|n_L|^2\approx \frac{4\pi}{2L+1}$.
By summing over all the rotational states and using
the completeness relation of the spherical harmonics, one obtains
\begin{align}
  &\sum_{LM}\left|\left<\Psi_{LM}^P\Psi^T_0 \right|e^{i\chi} \left|\Psi_0^P\Psi^T_0\right>\right|^2\notag\\
  &=\sum_{LM}\left| \frac{1}{\sqrt{4\pi}}\int d{\bm{\Omega}}\,Y_{LM}^*({\bm{\Omega}})
\mathcal{S}(\bm{b}; \bm{\Omega})
\left|\Phi^P(\bm{\Omega})\Psi^T_0\right>
\right|^2\\
&=\sum_{LM}\frac{1}{4\pi}\iint d{\bm{\Omega}} d{\bm{\Omega}}'\,
Y_{LM}^*({\bm{\Omega}})Y_{LM}({\bm{\Omega}}')
\mathcal{S}(\bm{b}; \bm{\Omega})
\mathcal{S}^\dagger(\bm{b}; \bm{\Omega}')
\\
  &=\frac{1}{4\pi}\int d{\bm{\Omega}}
\left|
\mathcal{S}(\bm{b}; \bm{\Omega})
\right|^2.
\end{align}
The above expression is particularly useful
because it allows the contributions of excited states
to be incorporated using only the intrinsic wave function.
One thus arrives at \cite{glauber1990}
\begin{align}
  \sigma_I\approx \int d\bm{b}\,\left(1-
\left<|\mathcal{S}(\bm{b}; \bm{\Omega})|^2\right>_{\bm{\Omega}}\right).
\label{icsaprx.eq}  
\end{align}
This formula is similar to that for the total reaction cross section,
Eq.~(\ref{rcsaprx.eq}), but differs in the treatment of the orientational
average. Equation~(\ref{icsaprx.eq}) corresponds to an incoherent average
over orientations, while Eq.~(\ref{rcsaprx.eq}) is obtained by performing
the average coherently at the amplitude level. 
See also Eqs. (47), (48), and (49) in Ref. \cite{Hagino2022}. 
Although Ref. \cite{Tostevin1999} interpreted Eq.~(\ref{icsaprx.eq})
as the total reaction cross section, such an identification
is not strictly valid. Instead, Eq.~(\ref{icsaprx.eq}) should be regarded
as an approximation to the interaction cross section discussed
in the present work.
The differential quasielastic scattering cross section
can also be derived as
\begin{align}
  \frac{d\sigma_{\rm qel}}{d\Omega_{\rm cm}}\approx
  \left<|f(\bm{q}; \bm{\Omega})|^2\right>_{\bm{\Omega}}.
\end{align}
A comparison with Eq.~(\ref{eldcsaprx.eq}) is also instructive
as they exhibit incoherent and coherent sums, respectively.

\subsection{Optical-limit approximations}
\label{ola.sec}

Although Eqs.~(\ref{elampaprx.eq}) and (\ref{rcsaprx.eq}) are simpler
than the original expression, its numerical evaluation still remains demanding.
To further reduce the computational cost, 
we introduce the optical-limit approximation. Depending upon how the integral for the orientation angles 
is treated, one can consider two different types of approximation. 

First, we consider a cumulant expansion
of the $S$-matrix before carrying out the angular averaging (Scheme 1).
Expressing
the phase shift function by the nucleon-nucleon profile function $\Gamma_{NN}$
\begin{align}
e^{i\chi(\bm{b})}\equiv\prod_{i\in P}\prod_{j\in T}[1-\Gamma_{NN}(\bm{b}+\hat{\bm{s}}^P_i-\hat{\bm{s}}_j^T)],
\end{align}
the leading order of the expansion reads
\begin{widetext}
\begin{align}
\mathcal{S}(\bm{b}; \bm{\Omega})
\approx
\exp\left[-\iint d\bm{r}^Pd\bm{r}^T \Gamma_{NN}(\bm{b}+\bm{s}^P-\bm{s}^T)
  \rho^P(\bm{r}^P;\bm{\Omega})\rho^T(\bm{r}^T)\right], 
\end{align}
\end{widetext}
with $\bm{r}=(\bm{s},z)$ and the one-body densities of the projectile and the target nuclei:
\begin{align}
  \rho^P(\bm{r};\bm{\Omega})&=\left<\Phi^P(\bm{\Omega})\right|\sum_{i\in P}\delta(\hat{\bm{r}}_i-\bm{r}) \left|\Phi^P(\bm{\Omega})\right>,\\
\rho^T(\bm{r})&=\left<\Psi^T_0\right|\sum_{i\in T}\delta(\hat{\bm{r}}_i-\bm{r}) \left|\Psi_0^T\right>.
\end{align}
It should be noted that the profile function is assumed to be spin independent, and the isospin labels are omitted for simplicity. If the isospin degrees of freedom are explicitly included, separate proton and neutron density distributions are required.
Although only the one-body density is involved at the present order,
the formalism can be systematically extended to higher orders.
In particular, the next-order term introduces the two-body density,
such as
\begin{align}
  \rho^P(\bm{r},\bm{r}';\bm{\Omega})&=\left<\Phi^P(\bm{\Omega})\right|\sum_{i<j\in P}\delta(\hat{\bm{r}}_i-\bm{r})\delta(\hat{\bm{r}}_j-\bm{r}') \left|\Phi^P(\bm{\Omega})\right>,
\end{align}
allowing higher-order correlations to be incorporated explicitly.
While the optical-limit approximation (OLA) provides an accurate description of nucleon-nucleus scattering, it is generally known to overestimate reaction cross sections in nucleus-nucleus collisions. As demonstrated in Ref.~\cite{Horiuchi26a,Horiuchi26b}, the inclusion of multiple-scattering contributions up to second order significantly improves the agreement with the full Glauber calculation.

Next, we consider expanding the $S$-matrix after performing the angular averaging (Scheme 2):
\begin{widetext}
\begin{align}
\frac{1}{4\pi}\int d{\bm{\Omega}}\,\mathcal{S}(\bm{b}; \bm{\Omega})
&\approx
\exp\left[-\frac{1}{4\pi}
\int d{\bm{\Omega}}\,\iint d\bm{r}^Pd\bm{r}^T \Gamma_{NN}(\bm{b}+\bm{s}^P-\bm{s}^T)
  \rho^P(\bm{r}^P;\bm{\Omega})\rho^T(\bm{r}^T)\right]\\
&=\exp\left[-\iint d\bm{r}^Pd\bm{r}^T \Gamma_{NN}(\bm{b}+\bm{s}^P-\bm{s}^T)
  \bar{\rho}^P(\bm{r}^P)\rho^T(\bm{r}^T)\right],
\end{align}
\end{widetext}
where
\begin{align}
\bar{\rho}^P(\bm{r}^P)\equiv \frac{1}{4\pi}\int d{\bm{\Omega}}\, \rho^P(\bm{r}^P;\bm{\Omega}).
\end{align}
This procedure is equivalent to the conventional approach~\cite{Horiuchi12},
where the density is averaged over all orientations prior to the calculation
and subsequently treated as a spherically symmetric density distribution.
Note that in Scheme 2
\begin{widetext}
\begin{align}
\frac{1}{4\pi}\int d{\bm{\Omega}}\,  \left|\mathcal{S}(\bm{b}; \bm{\Omega})\right|^2&=
  \exp\left\{\frac{1}{4\pi}\int d{\bm{\Omega}}\,\left[-\iint d\bm{r}^Pd\bm{r}^T \Gamma_{NN}(\bm{b}+\bm{s}^P-\bm{s}^T)
    \rho^P(\bm{r}^P;\bm{\Omega})\rho^T(\bm{r}^T)\right.\right.\notag\\
   & -\left.\left.\iint d\bm{r}^Pd\bm{r}^T \Gamma_{NN}^*(\bm{b}+\bm{s}^P-\bm{s}^T)
\rho^P(\bm{r}^P;\bm{\Omega})\rho^T(\bm{r}^T)\right]\right\}
  \\
&=\left|\exp\left[-\iint d\bm{r}^Pd\bm{r}^T \Gamma_{NN}(\bm{b}+\bm{s}^P-\bm{s}^T)
  \bar{\rho}^P(\bm{r}^P)\rho^T(\bm{r}^T)\right]\right|^2\\
&=\left|\frac{1}{4\pi}\int d{\bm{\Omega}}\,\mathcal{S}(\bm{b}; \bm{\Omega})\right|^2.
\end{align}
\end{widetext}
Therefore $\sigma_R=\sigma_I$ and $\frac{d\sigma_{\rm el}}{d\Omega_{\rm cm}}=\frac{d\sigma_{\rm qel}}{d\Omega_{\rm cm}}$.

\begin{widetext}
To facilitate comparison, we briefly summarize the expressions
for Schemes 1 and 2:

\subsubsection*{Scheme 1}

\paragraph*{Elastic scattering amplitude:}
\begin{align}
  f(\bm{q})=
\frac{1}{4\pi}
\int d{\bm{\Omega}}\,
f(\bm{q};{\bm{\Omega}})
\end{align}
with
\begin{align}
f(\bm{q};{\bm{\Omega}})=\frac{ik}{2\pi}\int d\bm{b}\,e^{-i\bm{q}\cdot\bm{b}}
\left\{1-
\exp\left[-\iint d\bm{r}^Pd\bm{r}^T \Gamma(\bm{b}+\bm{s}^P-\bm{s}^T)
  \rho^P(\bm{r}^P;\bm{\Omega})\rho^T(\bm{r}^T)\right]\right\}.
\end{align}

\paragraph*{Total reaction cross section:}

\begin{align}
\sigma_R=\int d\bm{b}\,\left(1-
\left|\frac{1}{4\pi}\int d\bm{\Omega}\,
\exp\left[-\iint d\bm{r}^Pd\bm{r}^T \Gamma(\bm{b}+\bm{s}^P-\bm{s}^T)
  \rho^P(\bm{r}^P;\bm{\Omega})\rho^T(\bm{r}^T)\right]
\right|^2
\right).
\end{align}

\paragraph*{`Interaction' cross section:}

\begin{align}
\sigma_I=\int d\bm{b}\,\left(1-
\frac{1}{4\pi}\int d\bm{\Omega}\,\left|
\exp\left[-\iint d\bm{r}^Pd\bm{r}^T \Gamma(\bm{b}+\bm{s}^P-\bm{s}^T)
  \rho^P(\bm{r}^P;\bm{\Omega})\rho^T(\bm{r}^T)\right]
\right|^2
\right).
\end{align}

\subsubsection*{Scheme 2}

\paragraph*{Elastic scattering amplitude:}

\begin{align}
f(\bm{q})=\frac{ik}{2\pi}\int d\bm{b}\,e^{-i\bm{q}\cdot\bm{b}}
\left\{1-
\exp\left[-\iint d\bm{r}^Pd\bm{r}^T \Gamma(\bm{b}+\bm{s}^P-\bm{s}^T)
  \bar{\rho}^P(\bm{r}^P)\rho^T(\bm{r}^T)\right]\right\}.
\end{align}

\paragraph*{Total reaction cross section:}

\begin{align}
\sigma_R=\sigma_I=\int d\bm{b}\,\left(1-
\left|
\exp\left[-\iint d\bm{r}^Pd\bm{r}^T \Gamma(\bm{b}+\bm{s}^P-\bm{s}^T)
  \bar{\rho}^P(\bm{r}^P)\rho^T(\bm{r}^T)\right]
\right|^2
\right).
\end{align}

\end{widetext}

Notice that the 
Scheme 1 requires the evaluation of the $S$-matrix
for each orientation of the intrinsic state,
whereas Scheme 2 requires only a single $S$-matrix calculation. 
If the angle dependence of observables is weak, one can expect that both schemes lead to 
similar results. On the other hand, if the angle dependence is strong, the two schemes yield 
considerably different results. This is the case for low-energy heavy-ion fusion reactions discussed 
in Sec. \ref{fusion.sec}. 

Although it is not central to the present discussion, it is worth briefly mentioning the nucleon-target profile-function (NTG) formalism~\cite{Ibrahim00R,Ibrahim00}, which has been widely used in Glauber-model calculations.
Whereas the conventional Glauber formalism treats the nucleon-nucleon interaction as the elementary process and performs the cumulant expansion accordingly, the NTG formalism regards the nucleon-target interaction as the elementary process. The nucleon-target profile function can be derived from the underlying nucleon-nucleon profile function. Although the leading-order NTG formalism requires the same input, namely the one-body density distributions, as the OLA, it provides an improved approximation beyond the simple OLA. Its effectiveness has been demonstrated in a number of applications~\cite{Horiuchi06, Horiuchi07, Takechi09, Horiuchi12,Nagahisa18}. Since the differences between the OLA and NTG formalisms are mainly quantitative and are not essential to the present discussion, we do not present NTG results in this work. Nevertheless, the corresponding expressions can be derived following the same procedure as described above, and are therefore summarized in Appendix~A.

\section{Results}
\label{results.sec}

\subsection{Model setup}
\label{model.sec}

Let us now numerically evaluate cross sections for intermediate-energy nuclear reactions of deformed 
nuclei.
As an input to the reaction calculations, one-body densities
of the projectile and target nuclei are required.
For this purpose, we adopt the standard two-parameter Fermi distribution
whose surface is deformed by an axially symmetric quadrupole deformation,
which is given by
\begin{align}
  \rho(\bm{r},\bm{\Omega})=\frac{\rho_0}{1+\exp\left[\left(r-R(\bm{\Omega},\bm{\Omega}_r)\right)/a\right]}, 
\end{align}
with $\bm{r}=(r,\bm{\Omega}_r)$ and
\begin{align}
  R(\bm{\Omega},\bm{\Omega}_r)=R_0\left(1+\beta_2\sqrt\frac{4\pi}{5}\sum_MY^*_{2M}(\bm{\Omega})Y_{2M}(\bm{\Omega}_r)\right).
\end{align}
In the present study, we investigate light, medium-heavy, and heavy deformed projectile nuclei
in a systematic manner, taking $^{24}$Mg, $^{56}$Fe, and $^{154}$Sm
as representative examples.
The density model contains three independent parameters.
They are determined according to the following procedure:
the diffuseness parameter $a$ is fixed with
the standard value $a=0.54$ fm \cite{BohrMottelson1}. 
The radius parameter $R_0$ is determined so as to reproduce
the experimental charge radius for each nucleus~\cite{Angeli13}.
The proton and neutron density distributions are assumed to be identical
apart from the normalization.

For the quadrupole deformation parameters, the empirical values are adopted, 
that is, $\beta_2$=0.601, 0.250, and 0.339 for $^{24}$Mg, $^{56}$Fe,
and $^{154}$Sm, respectively~\cite{Nudat3}.
These values are among the largest deformations
observed in their respective mass regions and thus provide
representative examples of strongly deformed nuclei.
To examine the sensitivity to the deformation, we also consider
an alternative density for $^{24}$Mg with a reduced deformation parameter
of $\beta_2=0.300$, which is approximately one half of the empirical value.

We consider both proton and carbon targets.
For the carbon target, the density distribution is assumed
to take a harmonic-oscillator form,
with its parameters determined so as
to reproduce the experimental charge radius.
The parameter set for the profile function is taken from Ref.~\cite{Ibrahim08}.
With these inputs fixed,
the calculation contains no further adjustable parameters.

\subsection{Cross sections}
\label{csres.sec}

\subsubsection{Elastic scattering}

\begin{figure*}[t]
  \centering
\includegraphics[width=0.9\linewidth]{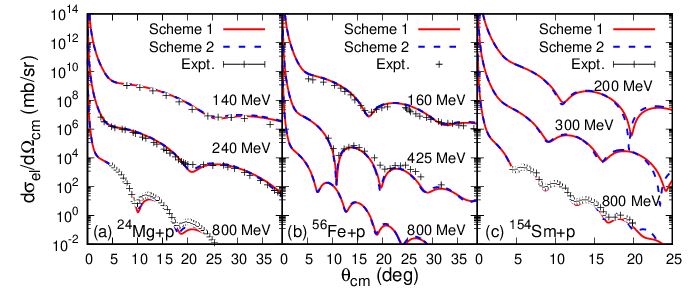}
\caption{Differential elastic scattering cross sections
  for $^{24}$Mg, $^{56}$Fe, and $^{154}$Sm on a proton target
  as a function of the scattering angles in the center-of-mass frame at various incident energies.
  For clarity, the first and second cross sections
  from the top are multiplied by factors of 10$^6$ and $10^3$, respectively.
  Experimental data are taken from Refs.~\cite{Barlett80,Hicks88,Schwandt82,DeLeo96,Blanpied88}.}  
\label{elastic-p.fig}
\end{figure*}

\begin{figure*}[t]
  \centering
\includegraphics[width=0.9\linewidth]{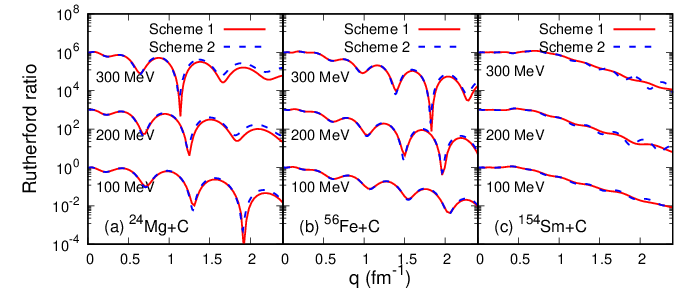}
\caption{Rutherford cross-section ratios 
  for $^{24}$Mg, $^{56}$Fe, and $^{154}$Sm on a carbon target as a function of the momentum transfer.  
  The first and second curves
  from the top are multiplied by factors of 10$^6$ and $10^3$, respectively.}
\label{elastic-C.fig}
\end{figure*}

Figure~\ref{elastic-p.fig} compares
the differential elastic scattering cross sections
for $^{24}$Mg, $^{56}$Fe, and $^{154}$Sm on a proton target
at various incident energies. 
These are plotted as a function of the scattering angles
in the center-of-mass frame.
The elastic Coulomb phase is incorpolated as prescribed in Ref.~\cite{Suzuki03}.
Experimental data are also shown for reference.
One can see that both Scheme 1 and Scheme 2 reproduce the measured cross sections
reasonably well over the entire angular range, including the backward angles,
indicating that the density distributions adopted in the present study
are sufficiently realistic. Furthermore, no significant difference
is observed between the two schemes. This is not surprising because,
as demonstrated in Refs.~\cite{Varga02,Nagahisa18,Horiuchi26a,Horiuchi26b},
the optical-limit approximation is known to work very well
for proton-nucleus scattering.

Figure~\ref{elastic-C.fig} shows
the Rutherford ratio of the differential elastic scattering cross sections
for those three projectiles on a carbon target as a function of
the momentum transfer.
While the two schemes yield almost identical results
at forward angles, a noticeable difference appears
at backward angles. However, because the Glauber theory
is not expected to provide a reliable description in this region,
the discrepancy has little impact on the overall picture.

\subsubsection{Total reaction cross sections}

\begin{figure}[thb]
  \centering
\includegraphics[width=1\linewidth]{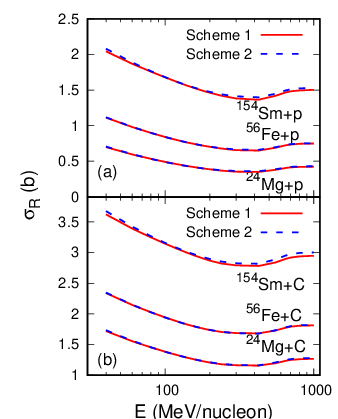}
\caption{Comparison of the total reaction cross sections calculated with
  Schemes 1 and 2 for $^{24}$Mg, $^{56}$Fe, and $^{154}$Sm on (a) proton and (b) carbon targets as
  a function of the incident energies.}
\label{rcs.fig}
\end{figure}

Figure~\ref{rcs.fig} compares the total reaction cross sections
over a wide range of incident energies. As anticipated from the behavior
of the elastic scattering cross sections, the difference between
the two schemes is rather small for $^{24}$Mg and $^{56}$Fe
on both proton and carbon targets, amounting to less than
at most approximately 10 mb.
These findings provide justification for extending the conventional Scheme 2,
which has primarily been employed for relatively light nuclei
to medium-heavy nuclear systems~\cite{Horiuchi12,Horiuchi22}.
The effect becomes much more pronounced for the heavier,
strongly deformed nucleus $^{154}$Sm, for which the difference reaches
approximately 30 mb for a proton target and 50 mb for a carbon target.
This can be understood
from the fact that deformation effects mainly originate from
the nuclear surface region; in heavier nuclei,
a larger number of nucleons contribute to the surface,
thereby amplifying the overall impact of deformation
on the total reaction cross section.

It is generally known that the optical-limit approximation
tends to overestimate the cross sections (see, e.g., Refs.~\cite{Horiuchi07,Horiuchi26b}).
Interestingly, the cross sections obtained with Scheme 1 
are systematically smaller than those of Scheme 2. 
This behavior can be understood as a consequence
of the multiple-scattering effects arising from intrinsic states
with different orientations, which are effectively incorporated in Scheme 1.

\subsubsection{Quasi-elastic scattering}

\begin{figure}[t]
  \centering
  \includegraphics[width=1\linewidth]{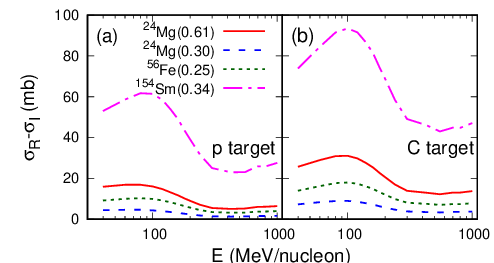}
\caption{Difference of the total reaction and the `interaction' cross sections
  as a function of incident energies for $^{24}$Mg, $^{56}$Fe, and $^{154}$Sm on (a) proton and (b) carbon targets. 
  The values in the parentheses indicate the values of the quadrupole deformation parameter 
  of the projectile nuclei adopted in each case.
  }
\label{rcsics.fig}
\end{figure}

Figure~\ref{rcsics.fig} shows the difference between
the total reaction and interaction cross sections as
a function of the incident energy. 
Only the results with Scheme 1 are presented,
since Scheme 2 yields $\sigma_R=\sigma_I$.
This quantity corresponds to
the sum of all inelastic scattering cross sections
leading to rotational excited states of the projectile nucleus.
For a proton target, the difference amounts to approximately 1--4\%
of the total reaction cross section, in agreement with the trend reported
in Ref.~\cite{Hatakeyama19}. For a carbon target, the difference is also modest,
remaining at the level of 1--3\% and thus comparable
to that for a proton target.
To examine the deformation dependence of the results,
we carried out an additional calculation for $^{24}$Mg
with the deformation parameter $\beta_2$ reduced by approximately
a factor of two. 
The results are shown by the dashed lines in the figure.
As expected from the $\beta_2$ dependence of
the transition strengths $\propto \beta_2^2$, the difference between
the total reaction and the interaction cross sections
is reduced by about a factor of four.
For $^{154}$Sm, the difference becomes considerably larger owing to its strong deformation. Nevertheless, the resulting magnitude remains within the expected range of order 100~mb predicted in Ref.~\cite{Kohama08}.

The effect of nuclear deformation on intermediate-energy nucleus-nucleus reactions 
can be more clearly illustrated by considering
the difference between the reaction probabilities obtained
with different orders of angular averaging, as defined in Eqs.~(\ref{rcsaprx.eq}) and (\ref{icsaprx.eq}):
\begin{align}
  \Delta^2(\bm{b})=
\left<\left|\mathcal{S}(\bm{b};\bm{\Omega})\right|^2\right>_{\bm{\Omega}}
-\left|\left<\mathcal{S}(\bm{b};\bm{\Omega})\right>_{\bm{\Omega}}\right|^2 
\label{variance.eq}
\end{align}
Obviously, $\int d\bm{b}\,\Delta^2(\bm{b})=\sigma_R-\sigma_I$.
This quantity is analogous to the variance of the reaction probability and
can be interpreted as the fluctuation of the reaction probability due to nuclear deformations.
Since the larger deformation generally leads to the larger density fluctuations
in the surface region,
one naturally expects this quantity to increase with deformation.

\begin{figure*}[htb]
  \centering
\includegraphics[width=0.8\linewidth]{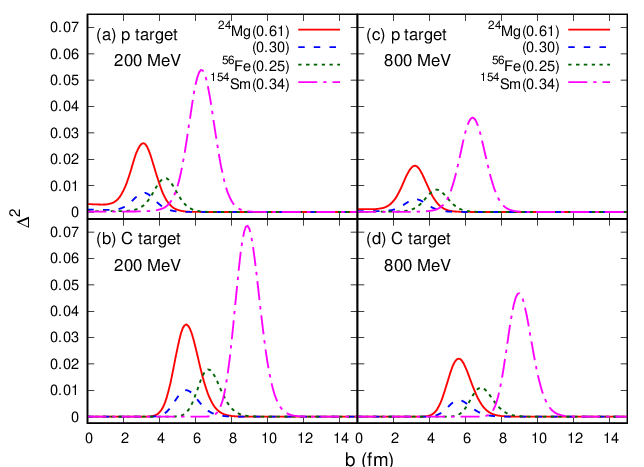}
\caption{The `varience' of the reaction probabilities defined by Eq. (\ref{variance.eq}) as a function of
  the impact parameter for $^{24}$Mg, $^{56}$Fe, and $^{154}$Sm on (a)(c) proton and (b)(d) carbon targets
at (a)(b) 200 and (c)(d) 800 MeV/nucleon. The values in the parentheses indicate the values of the quadrupole deformation parameter 
  of the projectile nuclei adopted in each case.}
\label{varience.fig}
\end{figure*}

Figure~\ref{varience.fig} displays the variances, $\Delta^2$,
at 200 and 800 MeV/nucleon on the proton and the carbon targets.
One can see that the fluctuation is largely concentrated in the nuclear surface region.
However, for a relatively light nucleus such as $^{24}$Mg,
it extends to smaller impact parameters as well.
This behavior arises from the fact that the absorption
is not yet complete in the interior region for such a light nucleus.
The degree of such incomplete absorption was quantified in Ref.~\cite{Makiguchi22}.
In the case of the proton target, the incident proton
can penetrate into the nucleus to some extent,
resulting in a broad distribution of the fluctuation over the impact parameter.
In contrast, for the carbon target, since the target is more absorptive,
the fluctuation is narrower and almost entirely localized
in the region where the projectile and the target nuclei overlap.

As the size of the projectile nucleus increases,
the peak of the fluctuation shifts toward larger impact parameters.
Consequently, a larger number of nucleons participate
in the relevant rotational excitations, which enhances
the contribution of inelastic channels.
As a result, the difference between the total reaction and
the interaction cross sections becomes larger for strongly deformed
heavier nuclei.
As the deformation of the projectile increases,
the overlap region where the two nuclei come into contact becomes
increasingly sensitive to orientation-dependent multiple-scattering effects
and inelastic contributions.
The observed trend therefore provides a natural manifestation
of the deformation effects in intermediate-energy nuclear reactions.

\section{Treatment of nuclear deformation in low-energy heavy-ion fusion reactions}
\label{fusion.sec}

Before we close the paper, it is instructive to discuss how nuclear deformation is treated in other 
nuclear reactions. For this purpose, we particularly discuss heavy-ion fusion reactions at sub-barrier 
energies~\cite{Hagino2012,Hagino2022}. 
For a deformed target, we assume the following Hamiltonian: 

\begin{equation}
    H=T+V(r,\theta_d)+H_{\rm rot}, 
\end{equation}

where $T=-\frac{\hbar^2}{2\mu}\vec{\nabla}^2$ with $\mu$ being the reduced mass 
is the kinetic energy for the relative motion between colliding nuclei, $V(r,\theta_d)$ is a deformed potential, and $H_{\rm rot}$ is the rotational Hamiltonian for the deformed target. Here, $\vec{r}$ is the relative 
coordinate between the two nuclei and 
$\theta_d$ is the orientation angle of the deformed target with respect to the direction of $\vec{r}$. 

We employ the iso-centrifugal approximation~\cite{Hagino2012,Hagino2022,hagino2004} and replace 
the centrifugal potential by $V_{\rm cent}(r)=J(J+1)\hbar^2/2\mu r^2$, where $J$ is the total angular momentum 
of the system. This considerably simplifies the angular momentum couplings for the coupled-channels calculations. 
In this approximation, the total wave function for the total angular momentum $J$ can be expanded as 
\begin{equation}
    \Psi_J(\vec{r},\theta_d)=\sum_I\frac{u^{(J)}_I(r)}{r}Y_{I0}(\theta_d), 
\end{equation}
where $I$ is the spin of the rotational state and the rotational wave function $Y_{I0}(\theta_d)$ 
satisfies $H_{\rm rot}|Y_{I0}\rangle=\epsilon_I|Y_{I0}\rangle$ with $\epsilon_I\propto I(I+1)$. 
Notice that in this approximation the angular momentum transfer is treated in an average way~\cite{hagino1995} 
and the angular momentum $I$ is used simply as a label of the channel wave functions. 
The coupled-channels equations for $u^{(J)}_I(r)$ are obtained from the Schr\"odinger equation, $H\Psi_J=E\Psi_J$, 
as
\begin{align}
    &\left(-\frac{\hbar^2}{2\mu}\frac{d^2}{dr^2}+\frac{J(J+1)^2\hbar^2}{2\mu r^2}+\epsilon_I-E\right)u^{(J)}_I(r)\notag\\
    &+\sum_{I'}\langle Y_{I0}|V(r,\theta_d)|Y_{I'0}\rangle u^{(J)}_{I'}(r)=0. 
    \label{cc:eq}
\end{align}
The $S$-matrix is obtained by solving these coupled-channels equations with scattering boundary conditions~\cite{Hagino2012,Hagino2022}, 
from which fusion cross sections can be constructed. 

When the rotational Hamiltonian $H_{\rm rot}$, thus the rotational energy $\epsilon_I$, can be neglected, 
the coupled-channels equations (\ref{cc:eq}) can be decoupled and fusion cross sections are computed as \cite{Hagino2012,Hagino2022,nagarajan1986}
\begin{equation}
\sigma_{\rm fus}(E)=\int^1_0d(\cos\theta_d)\sigma_{\rm fus}(E;V(r,\theta_d)),
\label{fusion-scheme1:eq}
\end{equation}
where $\sigma_{\rm fus}(E;V(r,\theta_d))$ is the fusion cross section calculated with a single-channel 
potential $\tilde{V}(r)=V(r,\theta_d)$ for a fixed value of $\theta_d$. 
This is nothing more than Scheme 1 discussed in the previous sections. That is, the fusion cross sections are 
firstly calculated for a fixed value of $\theta_d$ and then take the angle average afterwards. 
On the other hand, Scheme 2 corresponds to taking the angle average of the potential as 
\begin{equation}
V_{\rm eff}(r)=\int^1_0d(\cos\theta_d)\,V(r,\theta_d),
\label{fusion-scheme2:eq}
\end{equation}
and compute fusion cross sections with the effective potential, $V_{\rm eff}(r)$. 

\begin{figure}[htb]
  \centering
\includegraphics[width=\linewidth]{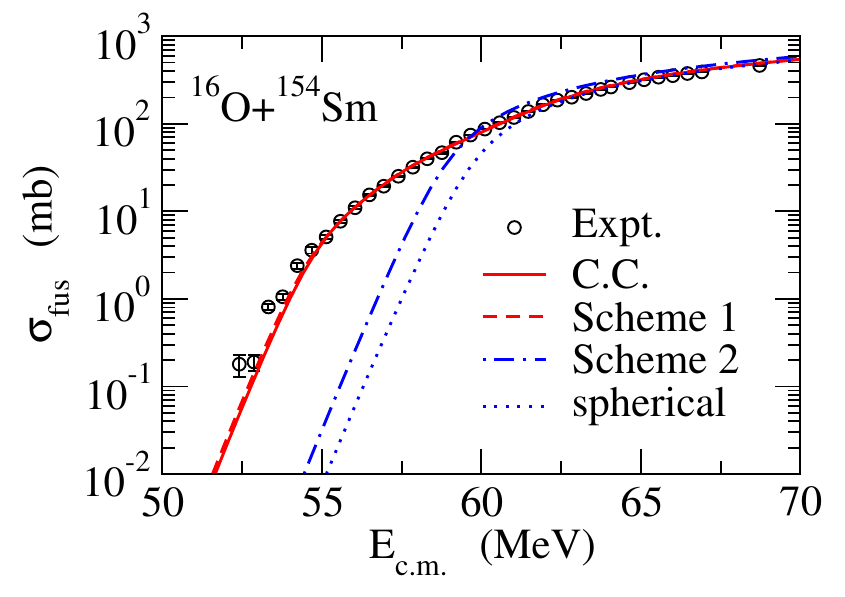}
\caption{
Fusion cross sections for the $^{16}$O+$^{154}$Sm system. 
The solid line denotes the results of the coupled-channels calculation, 
while the dashed and the dot-dashed lines are obtained in the 
adiabatic approximation with Scheme 1 [Eq. (\ref{fusion-scheme1:eq})] 
and Scheme 2 [Eq. (\ref{fusion-scheme2:eq})], respectively. 
The dotted line is obtained by assuming a spherical $^{154}$Sm. 
The experimental data are taken from Ref. \cite{Leigh1995}. 
}  
\label{fusion.fig}
\end{figure}

Figure~\ref{fusion.fig} compares Scheme 1 with Scheme 2 for the $^{16}$O+$^{154}$Sm system, together with the 
experimental data taken from Ref.~\cite{Leigh1995}. 
To this end, we use a deformed Woods-Saxon potential given by~\cite{ccfull}  
\begin{widetext}
\begin{eqnarray}
V(r,\theta_d)&=&V_N(r,\theta_d)+V_C(r,\theta_d), \\
V_N(r,\theta_d)&=&\frac{-V_0}{1+\exp[(r-R-R_T\beta_2Y_{20}(\theta_d)
-R_T\beta_4Y_{40}(\theta_d))/a]}, \\
V_C(r,\theta_d)&=&\frac{Z_PZ_Te^2}{r}+\left(\beta_2
+\frac{2}{7}\sqrt{\frac{5}{\pi}}\beta_2^2\right)
\,\frac{3Z_PZ_Te^2}{5}\frac{R_T^2}{r^3}
Y_{20}(\theta_d)
+\left(\beta_4
+\frac{9}{7}\beta_2^2\right)
\,\frac{3Z_PZ_Te^2}{9}\frac{R_T^4}{r^5}
Y_{40}(\theta_d), 
\end{eqnarray}
\end{widetext}
where $\beta_2$ and $\beta_4$ are the quadrupole and the hexadecapole deformation parameters of the deformed 
target nucleus, respectively, while $R_T$ is the radius of the target nucleus. 
$Z_P$ and $Z_T$ are the atomic numbers of the projectile and the target nuclei, respectively. 
We use $V_0$=165.0 MeV, $R=$7.486 fm, and $a$=1.05 fm. For the deformation parameters, we use 
$\beta_2=0.322$ and $\beta_4=0.05$ together with $R_T=5.682$ fm. The rotational excitation energies 
are set to be $\epsilon_I=0.0819I(I+1)/6$ MeV and we include up to 8$^+$ state in the ground state 
rotational band. Fusion is simulated with the incoming wave boundary condition~\cite{Hagino2012,Hagino2022,ccfull} by keeping the potential to be real. 

The solid line in the figure is obtained by solving the coupled-channels equations with the computer code {\tt CCFULL} \cite{ccfull}, while the dashed 
line is obtained using the adiabatic approximation with Scheme 1, Eq. (\ref{fusion-scheme1:eq}). 
One can see that the adiabatic approximation works well for this system. The small difference is due to 
the finite excitation energies of the rotational states: if $\epsilon_I$ are set to be zero, the two lines 
coincide with each other. On the other hand, the dot-dashed line denotes the result of Scheme 2 based on the 
effective potential given by Eq. (\ref{fusion-scheme2:eq}). Even though cross sections are enhanced as compared 
to the spherical case with $\beta_2=\beta_4=0$ denoted by the dotted line, the degree of enhancement is 
much smaller than the result of Scheme 1 and is inconsistent with the coupled-channels calculation. 
Evidently, Scheme 1 is a more appropriate choice compared to Scheme 2 in low-energy heavy-ion fusion reactions. 
Incidently, Scheme 2 systematically yields larger fusion cross sections at energies above the Coulomb barrier 
($E> 59.4$ MeV), which is consistent with Fig. \ref{rcs.fig}. 

\section{Comments on ultra-relativistic heavy-ion collisions}
\label{relhic.sec}

The formalism presented in this paper has a direct consequence to the Glauber approach~\cite{jia2022-2,ke2025,xu2026,miller2007,shou2015} as well as to hydrodynamical calculations 
for ultra-relativistic heavy-ion collisions. In the latter approach, the initial transverse entropy profile 
for quark gluon plasma (QGP) is generated using e.g. the \trento model~\cite{trento1,trento2}, which is then used 
as an input to hydrodynamic simulations. In the \trento model, the entropy profile is generated from deformed one-body densities with 
random orientations of deformed nuclei. A simulation is carried out on an event-by-event base, and final 
observables are constructed by taking an average over generated events. This corresponds to Scheme 1 discussed 
in this paper, that is, observables are first calculated with fixed orientations of deformed nuclei and then the 
orientation average is taken in order to compare with experimental data. 

Recently, Dobaczewski {\it et al.} questioned the approach of probing nuclear shapes using ultra-relativistic heavy-ion 
collisions \cite{doba2025}. Their argument is that the density of a deformed nucleus is spherical in the laboratory 
frame and the information of deformation cannot be deduced only with one-body densities, that is, two-body or higher-body densities are required to probe nuclear shape \cite{sun2026}. 

This is certainly true from the nuclear structure point of view. Indeed, if one takes Scheme 2 discussed in this paper, 
only spherical densities are involved in a calculation and it is hard to probe nuclear shape except for 
a modification of the densities due to deformation~\cite{Minomo11,Sumi12,Horiuchi12,Watanabe14,Horiuchi21,Horiuchi22}. 
However, one can take Scheme 1 to probe nuclear shape. As we emphasize, if observables have strong 
orientation dependence, one can probe nuclear shape from that dependence. 
We believe that this is the strategy of probing nuclear deformation with ultra-relativistic heavy-ion collisions. 
That is, nuclear shape is not directly probed, but only indirectly through reaction observables. 
We agree with Dobacewski {\it et al.} in a sense that it is not literally correct to say that nuclear shape can 
be directly determined with ultra-relativistic heavy-ion collisions. 

Another criticism of Dobaczewski {\it et al.} is that the nuclear rotation is a classical concept 
and it is not correct to compare the time scales~\cite{doba2025}. We agree with this. 
As in the case of low-energy heavy-ion fusion reactions discussed in the previous section, what really matters 
is the difference in the energy scale between the collision and the rotational degree of freedom. 
If the rotational energy is small as compared to the collision energy, one can safely introduce the 
adiabatic approximation and neglect the rotational Hamiltonian. Notice that this is equivalent to taking the needle approximation for the overlap between two deformed wave functions. On the other hand, it is often convenient 
and intuitive to introduce the concept of the time scale. This is the case also for low-energy heavy-ion 
reactions, analyzed with a semi-classical coupled-channels approach~\cite{Hagino2022,BrogliaWinther}. 
In this viewpoint, the energy of the rotational states can be interpreted 
in terms of the moment of inertia, which classically determines the time scale of the rotational motion~\cite{otsuka2026}. One should, however, bare in mind that it is the difference in the energy scale, rather than 
the difference in the time scale, which really matters in realistic quantum mechanical systems. 

\section{Conclusion}
\label{conclusion.sec}

In this work, we have derived a formalism that explicitly incorporates
the orientation effects of highly deformed intrinsic nuclear states
into the description of intermediate-energy nuclear scattering.
The formalism provides a transparent connection
between intrinsic deformation and scattering observables,
thereby clarifying how nuclear rotational motion manifests
itself in reaction processes at high energies.

We have further shown that an expansion
based on the optical-limit approximation
naturally contains contributions from a class of nuclear excitations.
This observation provides a new interpretation
of the conventional optical-limit treatment
and clarifies the physical content that is implicitly included
in the formalism. The magnitude of the effect remains modest
for light and medium-heavy nuclei,
supporting the validity of the conventional treatment adopted
in previous studies, while it is increasing appreciable for strongly deformed systems.
Exploiting this property,
a part of the inelastic scattering cross section can be estimated
in a simple and practical manner.
This, in turn, offers a useful framework
for evaluating experimentally relevant interaction cross sections,
which are among the most important observables in studies of unstable nuclei.
The present analysis also demonstrates
that the total inelastic scattering cross section summed over
all the rotationally excited states can be understood
as a manifestation of fluctuations
of the nuclear surface density associated with nuclear deformation.

The present work establishes the theoretical foundation for analyzing
deformation effects within the Glauber framework.
An important future direction is to perform comprehensive
and quantitative analyses
of total reaction and interaction cross sections
using more realistic microscopic wave functions.
Such studies will provide a deeper understanding
of the interplay between nuclear structure and reaction dynamics
and will contribute to more reliable extractions of nuclear radii
and density distributions from experimental data.

\begin{acknowledgments}
We thank the late Dr. Akihisa Kohama for useful discussions on the treatment of nuclear deformation in Glauber theory. He attempted to include the effect of nuclear deformation in the ``kurotama" model for intermediate-energy nuclear reactions and advocated Scheme 1 presented in this paper, even though the treatment was purely classical. Indeed, this work was originally inspired by his question regarding this problem. In addition, we also thank 
J.M. Yao, M. Kitazawa, G. Giacalone, T. Duguet, J. Jia for useful discussions during 
the workshop on ``Intersection of nuclear structure and high-energy nuclear collisions 2026" 
held during April 13-24, 2026, 
at Yukawa Institute for Theoretical Physics, Kyoto University.
This work was in part supported by 
  JSPS KAKENHI Grants No. 23K03414, No. 25K01005, No. 25K07285, No. 26H01416, No. 26K00703,  
and JSPS Bilateral Program No. JPJSBP120247715.  
\end{acknowledgments}

\appendix

\section{Nucleon-target formalism in the Glauber theory}

As discussed in the main text, the optical-limit approximation (OLA) retains only the leading-order contribution of the cumulant expansion. An alternative formulation, known as the nucleon-target profile-function (NTG) formalism~\cite{Ibrahim00R,Ibrahim00}, has been widely used to improve the description of nucleus-nucleus scattering within the Glauber theory. While the conventional Glauber formalism regards the nucleon-nucleon interaction as the elementary process, the NTG formalism introduces an effective nucleon-target profile function, which can be derived from the underlying nucleon-nucleon profile function. Since the NTG formalism requires the same one-body density distributions as input, its implementation closely parallels that of the OLA discussed in the main text. Although NTG results are not presented in this work because the differences are mainly quantitative and are not essential to the present discussion, we summarize the corresponding expressions here for completeness.
\begin{align}
\left<\Psi_0^P\Psi^T_0
\right|e^{i\chi}
\left|\Psi_0^P\Psi^T_0\right>
\to 
\left<\Psi_0^P
\right|\prod_{i\in P} e^{i\chi_{NT}(\bm{b}+\hat{\bm{s}}_i^P)}
\left|\Psi_0^P\right>
\end{align}
with 
\begin{align}
e^{i\chi_{NT}(\bm{b})}&\equiv 1-\Gamma_{NT}(\bm{b}),\\
&\equiv \left<\Psi_0^T
\right|\prod_{j\in T} \left[1-\Gamma_{NN}(\bm{b}-\hat{\bm{s}}_j^T)\right]
\left|\Psi_0^P\right>,
\label{psfNT.eq}
\end{align}
where $\Gamma_{NT}$ denotes the nucleon-target profile function.

Following the same approximation employed in the main text, the orientation-dependent $S$ matrix can be written as
\begin{align}
\mathcal{S}(\bm{b};\bm{\Omega})
&= \left<\Phi^P(\bm{\Omega})
\right|\prod_{i\in P} e^{i\chi_{NT}(\bm{b}+\hat{\bm{s}}_i^P)}
\left|\Phi^P(\bm{\Omega})\right>\\
&\approx \exp\left[-\int d\bm{r}^P\rho^P(\bm{r}^P;\bm{\Omega})\Gamma_{NT}(\bm{b}+\bm{s}^P)\right].
\label{OLAGNT1.eq}
\end{align}
Equation~(\ref{psfNT.eq}) can also be expanded as the same way
\begin{align}
1-\Gamma_{NT}(\bm{b})\approx \exp\left[-\int d\bm{r}^T\, \rho^T(\bm{r}^T)\Gamma_{NN}(\bm{b}-\bm{s}^T)\right].
\label{OLAGNT2.eq}
\end{align}
Substituting Eq.~(\ref{OLAGNT2.eq}) into Eq.~(\ref{OLAGNT1.eq}), we obtain the NTG expressions corresponding to Scheme~1 and Scheme~2:
\begin{widetext}
\begin{align}
\mathcal{S}(\bm{b};\bm{\Omega})\approx
\exp\left\{-\int d\bm{r}^P\rho^P(\bm{r}^P;\bm{\Omega})
\left[1-\exp\left(-\int d\bm{r}^T\, \rho^T(\bm{r}^T)\Gamma_{NN}(\bm{b}+\bm{s}^P-\bm{s}^T)\right)\right]\right\}
\end{align}
for Scheme 1 and
\begin{align}
\frac{1}{4\pi}\int d\bm{\Omega}\, \mathcal{S}(\bm{b};\bm{\Omega})\approx
\exp\left\{-\int d\bm{r}^P\,\bar{\rho}^P(\bm{r}^P)
\left[1-\exp\left(-\int d\bm{r}^T\, \rho^T(\bm{r}^T)\Gamma_{NN}(\bm{b}+\bm{s}^P-\bm{s}^T)\right)\right]\right\}
\end{align}
for Scheme 2, the latter of which corresponds to the conventional NTG formula~\cite{Ibrahim00R,Ibrahim00}.
\end{widetext}

\bibliography{ref}

\end{document}